\documentclass[journal]{IEEEtranTIE}
\usepackage{graphicx}
\usepackage{cite}
\usepackage{picinpar}
\usepackage{amsmath}
\usepackage{stfloats}
\usepackage{url}
\usepackage{flushend}
\usepackage[latin1]{inputenc}
\usepackage{colortbl}
\usepackage{soul}
\usepackage{multirow}
\usepackage{pifont}
\usepackage{color}
\usepackage{alltt}
\usepackage[hidelinks]{hyperref}
\usepackage{siunitx}
\usepackage{breakurl}
\usepackage{epstopdf}
\usepackage{pbox}
\usepackage{algorithm}
\usepackage{algpseudocode}
\usepackage{enumitem}
\usepackage{arydshln}
\makeatletter
\newcommand\footnoteref[1]{\protected@xdef\@thefnmark{\ref{#1}}\@footnotemark}

\usepackage{enumerate}
\makeatother

\begin{document}
\title{	Real-Time  Rejection and Mitigation of \\ Time Synchronization Attacks on the \\ Global Positioning System   }

\author{
	{
		Ali Khalajmehrabadi, \emph{Student Member, IEEE},
		Nikolaos Gatsis, \emph{Member, IEEE},
		\\ David Akopian, \emph{Senior Member, IEEE},
		and Ahmad F. Taha, \emph{Member, IEEE}
	}
	
	\thanks{
		
		{
			Authors are with the Electrical and Computer Engineering Department, University of Texas at San Antonio, San Antonio, TX,  78249, USA (e-mails: \{ali.khalajmehrabadi, nikolaos.gatsis, david.akopian, ahmad.taha\}@utsa.edu). 
		}
	}
}

\maketitle
	
\begin{abstract}
This paper introduces the Time Synchronization Attack Rejection and Mitigation (TSARM) technique for Time Synchronization Attacks (TSAs) over the Global Positioning System (GPS). The technique estimates the clock bias and drift of the GPS receiver along with the possible attack contrary to previous approaches. Having estimated  the time instants of the attack, the clock bias  and  drift of the receiver are corrected. The proposed technique is computationally efficient and can be easily implemented in real time, in a fashion complementary to standard algorithms for position, velocity, and time estimation in off-the-shelf receivers. The performance of this technique is evaluated on a set of collected data from a real GPS receiver. Our method renders excellent time recovery consistent with the application requirements. The numerical results demonstrate that the TSARM technique outperforms competing approaches in the literature.
\end{abstract}

\begin{IEEEkeywords}
Global Positioning System, Time Synchronization Attack, Spoofing Detection
\end{IEEEkeywords}

\definecolor{limegreen}{rgb}{0.2, 0.8, 0.2}
\definecolor{forestgreen}{rgb}{0.13, 0.55, 0.13}
\definecolor{greenhtml}{rgb}{0.0, 0.5, 0.0}

\vspace{-5pt}
\section{Introduction}
\label{Introduction}
\IEEEPARstart{I}{nfrastructures} such as  road tolling systems, terrestrial digital video broadcasting, cell phone and air traffic control towers,  real-time industrial control systems,  and  Phasor Measurement Units (PMUs)~\cite{ Energy} heavily rely on synchronized precise timing for consistent and accurate network communications to maintain records and ensure their traceability. The Global Positioning System (GPS) provides time reference of microsecond precision for these systems~\cite{GPSgov, r27, r1, Misra2006}. 
\par The GPS-based time-synchronization systems use the civilian GPS channels, which are open to the public~\cite{Shaked, Xu}. The unencrypted nature of these signals makes them vulnerable to unintentional interference and intentional attacks. Thus, unauthorized manipulation of GPS signals leads to disruption of correct readings of GPS-based time references, and thus, is called Time Synchronization Attack (TSA).  To address the impact of malicious attacks, for instance on PMU data, the Electric Power Research Institute  published a technical report that recognizes the vulnerability of PMUs to GPS spoofing under its scenario \textit{WAMPAC.12: GPS Time Signal Compromise}~\cite{NESCOR}. These attacks  introduce erroneous time stamps which are eventually equivalent to inducing wrong phase angle  in the PMU measurements~\cite{r15, CommSurvey}. The impact of TSAs on generator trip control, transmission line fault detection, voltage stability monitoring, disturbing event locationing, and power system state estimation has been studied and evaluated both experimentally \cite{r2} and through simulations~\cite{r12, r14, r13}.
\par Intentional unauthorized manipulation of GPS signals is commonly referred to as \textit{GPS spoofing}, and  can be categorized based on the spoofer mechanism as follows: 
\begin{itemize}[leftmargin=*]
\item \textit{Jamming (blocking)}: The spoofer sends high power signals to jam the normal operation of the receiver by disrupting the normal operation  of the  victim receiver, often referred to as \textit{loosing lock}. Then, the victim receiver may lock onto the spoofer signal after jamming \cite{r5,r6, r15, Todd}. 
\item \textit{Data level spoofing}: The spoofer manipulates  the navigation data such as orbital parameters (ephemerides) that are used to compute satellite locations~\cite{r14, r4,r5}.
\item \textit{Signal level spoofing}: The spoofer synthesizes  GPS-like signals that carry the same navigation data as concurrently  broadcasted by the satellites~\cite{r2}.
\item \textit{Record-and-replay attack}: The spoofer records the authentic GPS signals and retransmits them with selected delays at higher power~\cite{r3, r15}. Typically the spoofer starts from low power transmission and increases its power to force the receiver to lock onto the spoofed (delayed) signal. The spoofer may change  the transmitting signal properties such that the victim receiver miscalculates its estimates. 
\end{itemize}

  \begin{table*}[t!]
             \caption{GPS Spoofing Detection Techniques: Detection Domain and Implementation Aspects}
                               \label{table1}
                              \begin{center}
                              \begin{tabular} { |c|c|c|l|c| }
                                \hline
             \textbf{Method} &\textbf{ Attack Detection Domain}& \textbf{Attack} &\begin{tabular}[x]{@{}c@{}} \textbf{Implementation Aspects}\end{tabular}& \textbf{Relevant}\\
                              \hline
                                 \hline
                 EKF & GPS Navigation domain & Not Estimated &\begin{tabular}[x]{@{}l@{}} Benchmark for most common GPS receivers \end{tabular} &  Yes \\ 
                                \hline
                 CUSUM \cite{wirenet} & GPS baseband signal domain & Not Estimated & \begin{tabular}[x]{@{}l@{}}  Applies hypothesis testing on packets of received signal \end{tabular} & No \\ 
                                \hline
                  Ref. \cite{r24} & GPS baseband \& power grid domains & Not Estimated & \begin{tabular}[x]{@{}l@{}} Combines the statistics of carrier-to-noise ratio difference\\   between two GPS antennas   \end{tabular} & No \\ 
                                \hline
                 SPREE \cite{r17} & GPS baseband signal domain & Not Estimated & \begin{tabular}[x]{@{}l@{}}  Applies auxiliary peak tracking in the correlators of receiver  \end{tabular} & No\\ 
                                  \hline
                Ref. \cite{r8,r9} & GPS baseband signal domain & Not Estimated & \begin{tabular}[x]{@{}l@{}} Applies a position-information-aided vector tracking loop \\ \end{tabular} & No \\ 
                                                                    \hline
              Ref. \cite{r10,Jansen} & GPS navigation domain & Not Estimated &\begin{tabular}[x]{@{}l@{}}  Needs collaboration
               among multiple GPS receivers  \end{tabular} &  No\\ 
            \hline                               Ref. \cite{ParticleFilter} & GPS navigation domain & Not Estimated &\begin{tabular}[x]{@{}l@{}} Applies an anti-spoofing particle filter  \end{tabular} & Yes\\ 
                                   \hline
               Ref. \cite{r11} & GPS navigation domain & Not Estimated &\begin{tabular}[x]{@{}l@{}} Applies hypothesis testing on a GPS clock signature  \end{tabular} & Yes\\ 
                              
                               \hline
                TSARM & GPS navigation domain & Estimated &\begin{tabular}[x]{@{}l@{}} Applies a real-time optimization technique  \end{tabular} & - \\ 
                                              \hline
                              \end{tabular}
                              \end{center}
                               \end{table*}

Common off-the-shelf GPS receivers lack proper mechanisms to detect these attacks.  A group of studies have been directed towards evaluating the requirements for successful attacks, theoretically \cite{r6} and experimentally \cite{r2, Characterization, SpooferDesign, Performance}. For instance, the work in \cite{SpooferDesign}  has designed a real spoofer as a Software Defined Radio (SDR) that records authentic GPS signals and retransmits fake signals. It provides the option of manipulating various signal properties  for spoofing.
\vspace{-7pt}
\subsection{Spoofing Detection Techniques in the Literature}
\par The first level of countermeasures to reduce the effect of malicious attacks on GPS receivers typically relies on the Receiver Autonomous Integrity Monitoring (RAIM) \cite{r1}.   Off-the-shelf GPS receivers typically apply RAIM consistency checks to  detect the anomalies  exploiting measurement redundancies. For example, RAIM may evaluate the variance  of GPS solution residuals and consequently generate an alarm if it exceeds a predetermined threshold. Similar variance authentication techniques have been proposed in~\cite{r17, Teunissen2} based on hypothesis testing on the Kalman filter innovations; however, they are vulnerable to smarter attacks that pass RAIM checks or the innovation hypothesis testing. 
\par   A plethora of  countermeasures have been designed  to make the receivers robust against more sophisticated attacks \cite{r3,r4,r5,r15, r7,r8,r9, r10,r19,r20, r17, Todd, r24, ParticleFilter, r11}.  Vector tracking exploits the signals from all satellites jointly and feedbacks the predicted position, velocity, and time (PVT) to the internal lock loops \cite{r7,r8,r9}. If an attack occurs, the lock loops become unstable which is an indication of attack.  Cooperative GPS receivers can perfrom authentication check by analyzing the integrity of measurements through peer-to-peer communications \cite{r9,r10,r19,r20}.  Also, a quick sanity check  for stationary time synchronization devices is to monitor the estimated location. As the true location can be known \textit{a priori}, any large shift that exceeds the maximum allowable position estimation error can be an indication of attack \cite{r11}. The receiver carrier-to-noise receiver can be used as \textcolor{red}{an} indicator of spoofing attack~\cite{Todd}. In~\cite{r24}, the difference between the carrier-to-noise ratios of two GPS antennas has been proposed as a metric of PMU trustworthiness.  In addition, some approaches compare the receiver's clock behavior against its statistics in normal operation~\cite{r3, r7, r11}. 
\vspace{-7pt}
\subsection{Existing Literature Gaps}
\par As discussed above, prior research studies addressed a breadth of problems related to GPS spoofing. However, there are certain gaps that should still be addressed: 1) Most of the works do not provide analytical models for different types of spoofing attacks. The possible attacking procedure models are crucial for designing the countermeasures against the spoofing attacks. 2) Although some countermeasures might be effective for a certain type of attack, a comprehensive countermeasure development is still lacking for defending the GPS receiver. This is practically needed as the receiver cannot predict the type of attack. 3) The main effort in the literature is in detection of possible spoofing attacks. However, even with the spoofing detection, the GPS receiver cannot resume its normal operation, especially in PMU applications where the network's normal operation cannot be interrupted. So, the spoofing countermeasures should not only detect the attacks but also mitigate their effects so that the network can resume its normal operation. 4)  There is a need for simpler solutions which can be integrated with current systems. 
\vspace{-7pt}
\subsection{Contributions of This Work}
\par This work addresses the previously mentioned gaps for stationary time synchronization systems. To the best of our knowledge, this is the first work that provides the following major contributions: 1) The new method is not a mere spoofing detector; it also estimates the spoofing attack. 2) The spoofed signatures, i.e., clock bias and drift, are corrected using the estimated attack. 3) The new method detects the smartest attacks that maintain the consistency in the measurement set.
A descriptive comparison  between our solution and representative works in the literature is provided in Table \ref{table1}.  A review of the spoofing detection domain shows that most of the prior art operates at the baseband signal processing domain, which necessitates manipulation of the receiver circuitry. Hence, the approach in the present paper is compared only to those works whose detection methodology lies in navigation domain.
\par The proposed TSA detection and mitigation approach in this paper consists of two parts. First,  a dynamical model is introduced  which analytically models the attacks in the receiver's clock bias and drift.  Through a proposed novel  Time Synchronization Attack Rejection and Mitigation (TSARM) approach, the clock bias and drift are estimated along with the attack. Secondly, the estimated clock bias and drift are modified based on the estimated attacks so that the receiver would be able to continue its normal operation with corrected timing for the application.  The proposed method detects and mitigates the effects of the smartest and most consistent reported attacks  in which the position of the victim receiver is not altered  and  the attacks on the pseudoranges are consistent with the attacks on pseudorange rates.
\par Different from outlier detection approaches in \cite{RobustSmoother,farahmand2011doubly}, the proposed method detects the anomalous behavior of the spoofer even if the measurement integrity is preserved.  The spoofing mitigation scheme has the following desirable attributes: 1) It solves a small quadratic program, which makes it applicable to commonly used devices. 2) It can be easily integrated into existing systems without changing the receiver's circuitry or necessitating mulitple GPS receivers as opposed to~\cite{r7,r8, r9, r19, r17, r24}.  3) It can run in parallel with current systems and provide an alert if spoofing has occurred. 4) Without halting the normal operation of the system, corrected timing estimates can be computed. 
\par The proposed anti-spoofing technique has been evaluated  using a commercial GPS receiver with open-source measurements access \cite{GoogleSource}.  These measurements have been perturbed with  spoofing attacks specific to PMU operation.  Applying the proposed anti-spoofing technique shows that the clock bias of the receiver can be corrected  within the maximum allowable error in the PMU IEEE C37.118 standard~\cite{IEEEStandard}.
\par \emph{Paper Organization:} A brief description of the GPS is described in Section \ref{Problem Formulation}. Then, we provide the models for possible spoofing attacks in Section~\ref{Models of Time Synchronization Attacks}. Section  \ref{novel sparse attack  model and detection technique} elaborates on the proposed solution to detect and modify the effect of these attacks. Our solution is numerically evaluated in Section \ref{Numerical Results} followed by the conclusions in Section~\ref{Conclusion}.
\vspace{-5pt}
\section{GPS PVT Estimation} 
\label{Problem Formulation}
In this section,  a brief overview of the GPS Position, Velocity, and Time (PVT) estimation is presented.
\par The main idea of localization and timing through GPS is  trilateration, which relies on the known location of satellites as well as distance measurements between satellites and  the GPS receiver.  In particular, the GPS signal from satellite $n$ contains a set of navigation data, comprising the ephemeris and the almanac (typically updated every 2 hours and one week, respectively), together with the signal's time of transmission ($t_{n}$).  This data is used to compute the satellite's position $\mathbf{p}_{n}=[x_n(t_{n}),y_n(t_{n}),z_n(t_{n})]^T$ in Earth Centered Earth Fixed (ECEF) coordinates, through a function known to the GPS receiver. Let $t_R$ denote the time that the signal arrives at the GPS receiver. The distance between the user (GPS receiver) and  satellite $n$ can be found by multiplying the signal propagation time $t_R-t_{n}$ by the speed of light $c$. This quantity is called \textit{pseudorange}:
$
\rho_{n}=c(t_{R}-t_{n}), \ n=1,\ldots,N
$, where $N$ is the number of visible satellites. The pseudorange is not the exact distance because the receiver and satellite clocks are both biased  with respect to the absolute GPS time. Let the receiver and satellite clock biases be denoted by $b_\mathrm{u}$ and $b_n$, respectively. Therefore,  the time of reception $t_R$ and $t_{n}$ are related to their absolute values in GPS time as follows:
$
 t_{R}=t_{R}^{\mathrm{GPS}}+b_{\mathrm{u}}; \quad
 t_{n}=t_{n}^{\mathrm{GPS}}+b_{n}, \; n=1,\ldots,N.
$
The $b_n$'s are computed from the received navigation data and are considered known. However, the bias $b_\mathrm{u}$ must be estimated and should be subtracted from the measured $t_R$ to yield the receiver absolute  GPS time $t_{R}^{\mathrm{GPS}}$, which can be used as a time reference used for synchronization. Synchronization systems time stamp their readings based on the Coordinated Universal Time (UTC) which has a known offset with the GPS time as
$
t_R^{\mathrm{UTC}}=t_R^{\mathrm{GPS}}-\Delta t_{ \mathrm{UTC}}
$, where $\Delta t_{ \mathrm{UTC}}$ is available online.\footnote{\url{https://confluence.qps.nl/qinsy/en/utc-to-gps-time-correction- \\ 32245263.html} (accessed Jan. 16, 2018).}
\par Let $\mathbf{p}_{\mathrm{u}}=[x_{\mathrm{u}},y_{\mathrm{u}},z_{\mathrm{u}}]^T$ be the coordinates of the GPS receiver, and $d_{n}$ its true range to satellite $n$. This distance is expressed via the locations $\mathbf{p}_\mathrm{u}$, $\mathbf{p}_n$ and the times $t_R^{\mathrm{GPS}}$,  $t_{n}^{\mathrm{GPS}}$ as $d_n=\|\mathbf{p}_n-\mathbf{p}_{\mathrm{u}}\|_2=c(t_R^{\mathrm{GPS}}-t_{n}^{\mathrm{GPS}})$. Therefore, the measurement equation becomes 
\vspace{-5pt}
\begin{equation}
\label{eq1}
\begin{split}
\rho_{n}=\Arrowvert \mathbf{p}_{n}- \mathbf{p}_{\mathrm{u}} \Arrowvert_2+c(b_{\mathrm{u}}-b_{n})+\epsilon_{\rho_{n}}
\end{split}
\end{equation}
where $ n =1, \ldots, N$, and $\epsilon_{\rho_{n}}$ represents the noise.
The unknowns in~\eqref{eq1} are $x_{\mathrm{u}},y_{\mathrm{u}},z_{\mathrm{u}},b_\mathrm{u}$ and therefore measurements from at least four satellites are needed to estimate them. 
\par Furthermore, the nominal carrier frequency ($ f_{c}=1575.42 \ \mathrm{MHz}$) of the transmitted signals from the satellite experiences a Doppler shift at the receiver due to the relative motion between the receiver and the satellite. Hence, in addition to pseudoranges, pseudorange rates are estimated from the Doppler shift and are related to the relative satellite velocity $\mathbf{v}_{n}$ and the user velocity  $\mathbf{v}_{\mathrm{u}}$ via
\vspace{-5pt}
 \begin{equation}
 \label{eq24}
 \begin{split}
 \dot{\rho}_{n}  = (\mathbf{v}_{n}-\mathbf{v}_{\mathrm{u}})^{T}\frac{\mathbf{p}_{n}-\mathbf{p}_{\mathrm{u}}}{\Arrowvert \mathbf{p}_{n}-\mathbf{p}_{\mathrm{u}} \Arrowvert}+\dot{b}_{\mathrm{u}}
 +\epsilon_{\dot{\rho}_{n}}
 \end{split}
 \end{equation}
where $\dot{b}_{u}$ is the clock drift. 
\par In most cases, there are more than four visible satellites, resulting in an overdetermined system of equations in~\eqref{eq1} and~\eqref{eq24}.  Typical GPS receivers use nonlinear Weighted Least Squares (WLS) to solve~\eqref{eq1} and~\eqref{eq24} and provide an estimate of the location, velocity, clock bias, and clock drift of the receiver, often referred to as PVT solution. To additionally exploit the consecutive nature of the estimates, a dynamical model is used.  The conventional dynamical model  for stationary receivers is a random walk model  \cite[Chap. 9]{r27} 
\vspace{-5pt}
\begin{equation}
\label{randomwalkmodel}
\begin{pmatrix}
x_{\mathrm{u}} [l+1] \\ y_{\mathrm{u}}[l+1]  \\ z_{\mathrm{u}}[l+1]  \\ b_{\mathrm{u}}[l+1]  \\ \dot{b}_{\mathrm{u}}[l+1] 
\end{pmatrix} = \begin{pmatrix}
\begin{array}{c;{2pt/2pt}r}
    \mathbf{I}_{3 \times 3} & \begin{matrix} \mathbf{0}_{3 \times 2} \end{matrix}  \\ \hdashline[1.5pt/1.5pt]
    \mathbf{0}_{2 \times 3} & \begin{matrix} 1 & \Delta t \\ 0 & 1 \end{matrix}  
    \end{array}
\end{pmatrix} \begin{pmatrix}
x_{\mathrm{u}} [l] \\ y_{\mathrm{u}}[l]  \\ z_{\mathrm{u}}[l]  \\ b_{\mathrm{u}}[l]  \\ \dot{b}_{\mathrm{u}}[l] 
\end{pmatrix}+ \mathbf{w}[l]
\end{equation}
where  $l$ is the time index,  $\Delta t$ is the time resolution (typically 1 sec), and $\mathbf{w}$ is the noise.  The dynamical system \eqref{randomwalkmodel} and  measurement equations \eqref{eq1} and \eqref{eq24} are the basis for estimating the user PVT using the Extended Kalman Filter (EKF). 
\par  Previous works have shown that simple attacks are able to mislead the solutions of WLS or EKF. Stationary GPS-based time synchronization systems  are currently equipped with the position-hold mode option which can potentially detect an attack if the GPS position differs from a known receiver location by a maximum allowed error~\cite{Arbiter}.  This  can be used as the first indication of attack. But, more advanced spoofers, such as the ones developed in~\cite{SpooferDesign}, have the ability to manipulate the clock bias and drift estimates of the stationary receiver without altering its position and velocity (the latter should be zero). So, even with EKF on the conventional dynamical models, perturbations on the pseudoranges in \eqref{eq1} and pseudorange rates in \eqref{eq24} can be designed so that they directly result in  clock bias and drift perturbations without altering the position and velocity of the receiver.
\vspace{-5pt}
\section{ Modeling Time Synchronization Attacks}
\label{Models of Time Synchronization Attacks}
This section puts forth a general attack model that encompasses the attack types discussed in the literature. This model is instrumental for designing the anti-spoofing technique discussed in the next section. 
\par While TSAs have different physical mechanisms, they manifest themselves as attacks on pseudorange and pseudorange rates. These attacks can be modeled as direct perturbations on \eqref{eq1} and \eqref{eq24} as
\vspace{-5pt}
\begin{equation}
\label{attack}
\begin{split}
\rho_{s}[l]=\rho [l]+s_{\rho}[l]\\
\dot{\rho}_{s}[l]=\dot{\rho} [l]+s_{\dot{\rho}}[l]
\end{split}
\end{equation}
where $s_{\rho}$ and $s_{\dot{\rho}}$ are the spoofing perturbations on pseudoranges and pseudorange rates, respectively; and $\rho_{s}$ and $\dot{\rho}_{s}$ are respectively the spoofed pseudorange and pseudorange rates.
\par A typical spoofer follows practical considerations to introduce feasible attacks. These considerations can be formulated as follows: 1) An attack is meaningful if it infringes the maximum allowed error defined in the system specification. {For instance in PMU applications, the attack should exceed the maximum allowable error tolerance specified by the IEEE C37.118 Standard, which is $1 \%$ Total Variation Error (TVE), equivalently expressed as $0.573^{\circ}$ phase angle error, $26.65 \ \mu \mathrm{s}$ clock bias error, or $\mathrm{7989} \ \mathrm{m}$ of distance-equivalent bias error \cite{IEEEStandard}. On the other hand, CDMA cellular networks require timing accuracy of 10 $\mu$s.\footnote{\url{http://www.endruntechnologies.com/cdma} (accessed Sept. 11, 2017).}  2) Due to the  peculiarities of the GPS receivers, the internal feedback loops may loose lock on the spoofed signal if the spoofer's signal properties change rapidly \cite{Characterization, r2}.  3)  The designed  spoofers have  the ability to manipulate the clock drift (by manipulating  the Doppler frequency) and clock bias (by manipulating the code delay)~\cite{SpooferDesign}. These  perturbations can be applied separately, however, the smartest attacks maintain the consistency of the spoofer's transmitted signal. This means that the pertubations on pseudoranges, $s_{\rho}$, are the integration of perturbations over pseudorange rates,  $s_{\dot{\rho}}$, in \eqref{attack}.

%\begin{figure}
%\begin{tabular}{ll}
%\includegraphics[scale=0.22]{AttackII}
%&
%\includegraphics[scale=0.235]{AttacKI}
%\end{tabular}
%\caption{ Type I attack (left) and Type II attack (right) on pseudorange and pseudorange rates. }
%\label{Spoofer}
%\end{figure}

\begin{figure}
	\centering \includegraphics[scale=0.45]{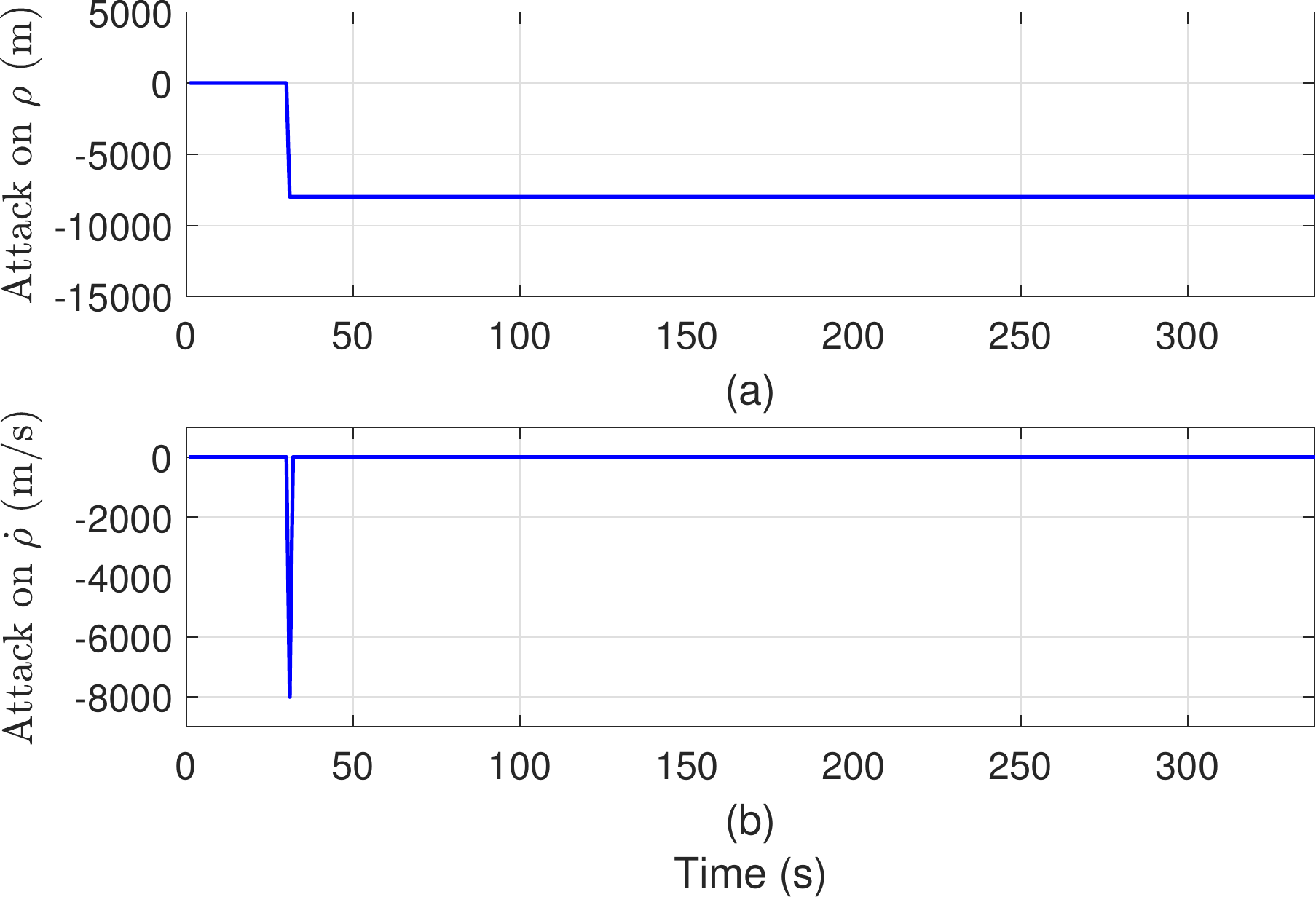}
	\caption{Type I attack  on (a) pseudorange and (b) pseudorange rate versus local observation time.}
	\label{SpooferI}
\end{figure}
\begin{figure}
	\centering \includegraphics[scale=0.45]{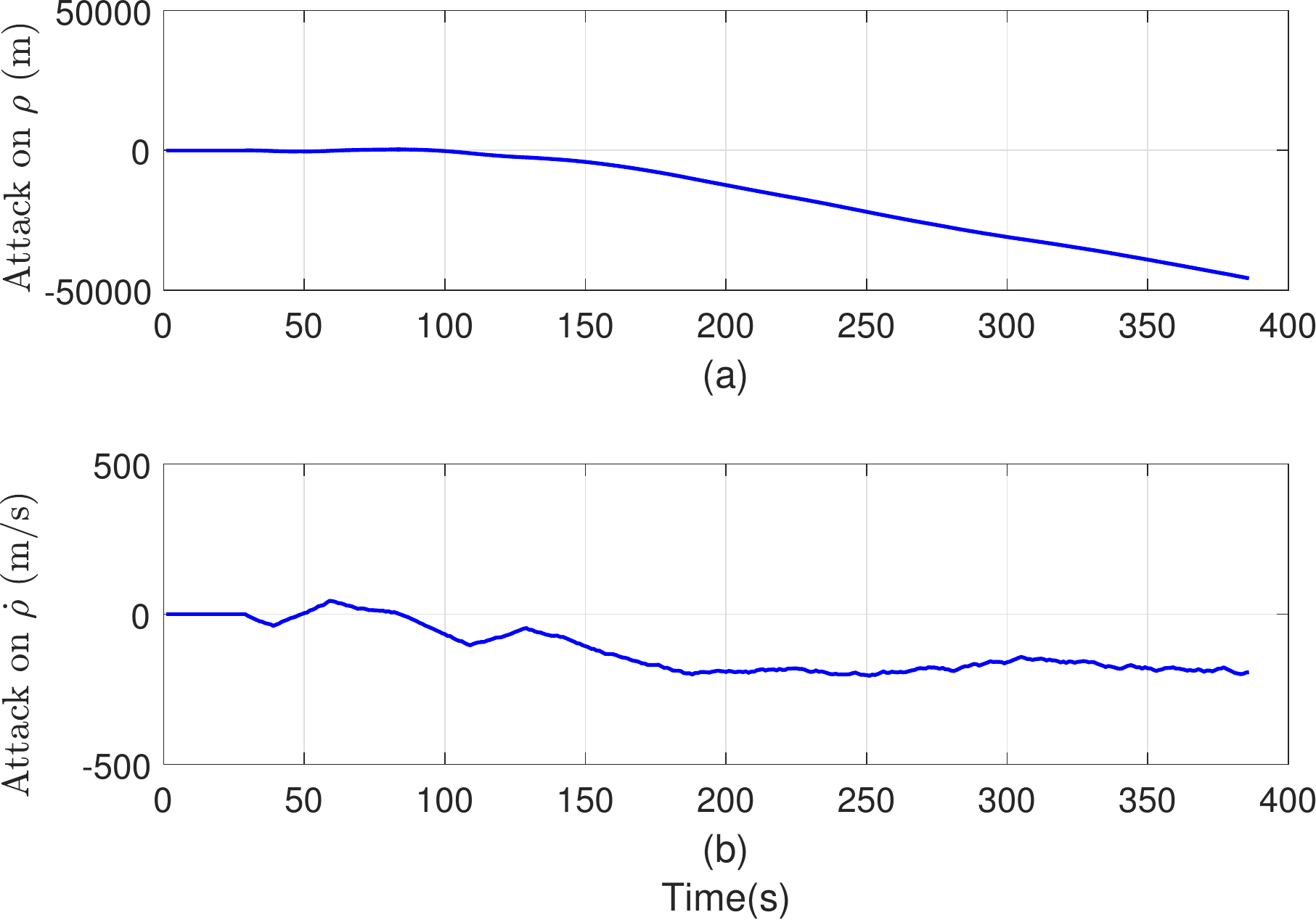}
	\caption{Type II attack  on (a) pseudorange and (b) pseudorange rate versus local observation time. }
	\label{SpooferII}
\end{figure}
\par Here, distinguishing between two attack procedures is advantageous as the literature includes very few research reports  on the technical intricacies of the spoofer constraints: 
\begin{itemize}[leftmargin=*]
\item Type I: The spoofer manipulates the authentic signal so that the bias abruptly changes in a very short time \cite{r14,r11, r5}. %Assuming that the spoofer does not jam the receiver {\color{red} [why do we need to say this here? I brought up here to emphasize that we do not model the jamming. If jamming happens, the early times of the attack look like a noise. We can remove if you think it is not needed.]} (which can be an indication of attack), 
Fig. \ref{SpooferI} illustrates this attack. The attack on the pseudoranges suddenly appears  at $t=\mathrm{30 \ s}$ and perturbs the pseudoranges by $\mathrm{8000} \ \mathrm{m}$. The equivalent attack on pseudorange rates is a Dirac delta function.

\item Type II: The spoofer gradually manipulates  the authentic signals and changes the clock bias through time~\cite{r2, Characterization,  r11, 6400273, Todd, r3}. This attack can be modeled by
\vspace{-5pt}
\begin{equation}
\label{Attackmodel}
\begin{split}
s_{\rho}[l]=s_{\rho}[l-1]+s_{\dot{\rho}}[l] \Delta t \\
s_{\dot{\rho}}[l]=s_{\dot{\rho}}[l-1]+\dot{s}_{\dot{\rho}}[l] \Delta t
\end{split}
\end{equation}
where $s_{\dot{\rho}}$ and $\dot{s}_{\dot{\rho}}$ are respectively called distance equivalent velocity and distance equivalent acceleration of the attack. To maintain the victim receiver lock on the spoofer's signals, the attack should not exceed a certain distance equivalent velocity. Two such limiting numbers are reported in the literature, namely, $\arrowvert s_{\dot{\rho}} \arrowvert \le \mathrm{400 \ m/s}$ in \cite{Characterization} and  $\arrowvert s_{\dot{\rho}} \arrowvert \le \mathrm{1000 \ m/s}$ in \cite{r2}. The acceleration to reach the maximum spoofing velocity is reported to be  $ \arrowvert \dot{s}_{\dot{\rho}} \arrowvert  \le \mathrm{5 \ m/s^{2}}$. The spoofer acceleration $\dot{s}_{\dot{\rho}}$ can be random, which makes Type II attack quite general. The distance equivalent velocity can be converted to the equivalent bias change rate (in $\mathrm{s}/\mathrm{s}$) through dividing the velocity by the speed of light. Fig.~\ref{SpooferII}  illustrates this attack. The attack on the pseudoranges starts at $t=\mathrm{30 \ s}$ and perturbs the pseudoranges gradually with distance equivalent velocity not exceeding $\mathrm{400 \ m/s}$  and maximum distance equivalent  random acceleration satisfying $\arrowvert \dot{s}_{\dot{\rho}} \arrowvert \le \mathrm{5 \ m/s^{2}}$. 
\end{itemize}

\par The introduced attack models are quite general and can mathematically capture most attacks on the victim receiver's measurements (pseudoranges and pseudorange rates) discussed in Section I. In another words, Type I and Type II attacks can be the result of data level spoofing, signal level spoofing, record-and-replay attack, or a combination of the aformentioned attacks. The main difference between Type I and Type II attacks is the spoofing speed. The speed of the attack depends on the capabilities of the spoofer with respect to manipulating various features of the GPS signals. Indeed, attacks of different speeds have been reported in the literature provided earlier in the present section. This work does not deal with jamming, which disrupts the navigation functionality completely whereas spoofing misleads it.

\par In the next section, a dynamical model for the clock bias and drift is introduced which incorporates these attacks. Based on this dynamical model,  an optimization problem to estimate these attacks along with the clock bias and drift is proposed.
\vspace{-5pt} 
\section{TSA-Aware Dynamical Model, TSA Rejection and Mitigation}
\label{novel sparse attack  model and detection technique}
This section introduces a dynamical model to accommodate the spoofing attack and a method to estimate the  attack. Afterwards, a procedure for approximately nullifing the effects of the attack on the clock bias and drift is introduced.
\vspace{-5pt}
\subsection{Novel TSA-aware Dynamical Model}
Modeling of the attack on  pseudoranges and pseudorange rates is motivated by the attack types discussed in the previous section. These attacks do not alter the position or velocity, but only the clock bias and clock drift. Our  model does not follow the conventional dynamical model  for stationary receivers which  allows the position of the receiver to follow a random walk model \eqref{randomwalkmodel}. Instead, the \emph{known} position and velocity of the victim receiver are exploited jointly. The state vector contains the clock bias and clock drift, and the attacks  are explicitly modeled on these components, leading to the following dynamical model:
\begin{equation}
\label{statemodel}
\small
\begin{split}
& \underbrace{\begin{pmatrix}
cb_{\mathrm{u}}[l+1]\\c\dot{b}_{\mathrm{u}}[l+1]
\end{pmatrix}}_{ \textstyle \mathbf{x}_{l+1}}= \underbrace{\begin{pmatrix}
1 & \Delta t\\ 0 & 1
\end{pmatrix}}_{ \textstyle \mathbf{F}}\underbrace{\begin{pmatrix}
cb_{\mathrm{u}}[l]\\ c\dot{b}_{\mathrm{u}}[l]
\end{pmatrix}}_{ \textstyle \mathbf{x}_{l}}+
\underbrace{\begin{pmatrix}
cs_{b}[l]\\ cs_{\dot{b}}[l]
\end{pmatrix}}_{ \textstyle \mathbf{s}_{l}}
+\underbrace{\begin{pmatrix}
c w_{b}[l]\\ c w_{\dot{b}}[l]
\end{pmatrix}}_{ \textstyle \mathbf{w}_{l}}
\end{split}
\end{equation}
where $s_{b}$ and $s_{\dot{b}}$ are the attacks on clock bias and clock drift and $w_{b}$ and $w_{\dot{b}}$ are colored Gaussian noise samples with covariance function defined in \cite[Chap. 9]{r27}. Here, both sides are multiplied with $c$, which is a typically adopted convention.  The state noise covariance matrix, $\mathbf{Q}_{l}$, is particular to the crystal oscillator of the device.
\par Similarly, define $\boldsymbol{\rho}[l]=[\rho_{1}[l],\ldots,\rho_{N}[l]]^{T}$ and $\boldsymbol{\dot{\rho}}[l]=[\dot{\rho}_{1}[l],\ldots,\dot{\rho}_{N}[l]]^{T}$. The measurement  equation can be as
\vspace{-5pt}
\begin{equation}
\label{measurementmodel}
\small
\begin{split}
& \underbrace{\begin{pmatrix}
\boldsymbol{\rho}[l]\\ \boldsymbol{\dot{\rho}}[l]
\end{pmatrix}}_{\textstyle \mathbf{y}_{l}}=\underbrace{\begin{pmatrix}
 \mathbf{1}_{N \times 1}&\mathbf{0}_{N \times 1}\\\mathbf{0}_{N \times 1} & \mathbf{1}_{N \times 1}
\end{pmatrix}}_{\textstyle \mathbf{H}}\underbrace{\begin{pmatrix}
cb_{\mathrm{u}}[l]\\ c\dot{b}_{\mathrm{u}} [l]
\end{pmatrix}}_{\textstyle \mathbf{x}_{l}}+ \\ 
& \underbrace{\begin{pmatrix}
\Arrowvert \mathbf{p}_{1}[l]-\mathbf{p}_{\mathrm{u}}[l]\Arrowvert\\\vdots\\\Arrowvert \mathbf{p}_{N}[l]-\mathbf{p}_{\mathrm{u}}[l]\Arrowvert\\
 (\mathbf{v}_{1}[l]-\mathbf{v}_{\mathrm{u}}[l])^{T}.\frac{\mathbf{p}_{1}[l]-\mathbf{p}_{\mathrm{u}}[l]}{\Arrowvert \mathbf{p}_{1}[l]-\mathbf{p}_{\mathrm{u}}[l] \Arrowvert}\\ \vdots \\ (\mathbf{v}_{N}[l]-\mathbf{v}_{\mathrm{u}}[l])^{T}.\frac{\mathbf{p}_{N}[l]-\mathbf{p}_{\mathrm{u}}[l]}{\Arrowvert \mathbf{p}_{N}[l]-\mathbf{p}_{\mathrm{u}}[l] \Arrowvert}
\end{pmatrix}-\begin{pmatrix}
cb_{1}[l] \\ \vdots \\ cb_{N}[l] \\ c\dot{b}_{1}[l] \\ \vdots \\ c\dot{b}_{N}[l]
\end{pmatrix} }_{\textstyle \mathbf{c}_{l}} + \underbrace{\begin{pmatrix}
\epsilon_{\boldsymbol{\rho}_{1}} [l]\\ \vdots \\ \epsilon_{\boldsymbol{\rho}_{N}}[l] \\ \epsilon_{\dot{\boldsymbol{\rho}}_{1}}[l] \\ \vdots \\ \epsilon_{\dot{\boldsymbol{\rho}}_{N}} [l]
\end{pmatrix}}_{ \textstyle \boldsymbol \epsilon_{l}}.
\end{split}
\end{equation}

Explicit modeling of $\mathbf{p}_{\mathrm{u}}$ and $\mathbf{v}_{\mathrm{u}}$ in $\mathbf{c}_{l}$ indicates that the dynamical model benefits from  using the stationary victim receiver's known position and velocity (the latter is zero).  The measurement noise covariance matrix, $\mathbf{R}_{l}$, is obtained through the measurements in the receiver. Detailed explanation of how to obtain the state and measurement covariance matrices, $\mathbf{Q}_{l}$ and $\mathbf{R}_{l}$, is provided in Section \ref{Numerical Results}.  It should be noted that the state covariance $\mathbf{Q}_{l}$  only depends on the victim receiver's clock behavior and does not change under spoofing. However, the measurement covariance matrix,  $\mathbf{R}_{l}$, experiences contraction.   The reason is that to ensure that the victim receiver maintains lock to the fake signals, the spoofer typically applies a power advantage over the real incoming GPS signals at the victim receiver's front end~\cite{Todd}. 
\par Comparing \eqref{Attackmodel}, \eqref{statemodel} and \eqref{measurementmodel}, TSAs which do not alter the position and velocity transfer the attack on pseudoranges and pseudorange rates directly  to clock bias and clock drift. Thus, it holds that $s_{\dot{\rho}}=cs_{b}$ and $\dot{s}_{\dot{\rho}}=cs_{\dot{b}}$. 
\vspace{-5pt}
\subsection{Attack Detection}
 Let $l=k, \ldots, k+L-1$ define the time index within the observation window of length $L$, where $k$ is the running time index. The solution to the dynamical model of \eqref{statemodel} and \eqref{measurementmodel} is obtained through stacking $L$ measurements and forming the following optimization problem:
 \vspace{-5pt}
\begin{equation}
\label{proposedapproach}
\begin{split}
(\hat{\mathbf{x}},\hat{\mathbf{s}})& =  \underset{\mathbf{x},\mathbf{s}}{\text{argmin}}\left\lbrace  \frac{1}{2} \sum_{l=k}^{k+L-1} \Arrowvert \mathbf{y}_{l}-\mathbf{H}\mathbf{x}_{l} -\mathbf{c}_{l}\Arrowvert_{\mathbf{R}_{l}^{-1}}^{2}  \right.  \\
 & \mspace{-50mu} \left.  +\: \frac{1}{2}\sum_{l=k}^{k+L-1}\Arrowvert \mathbf{x}_{l+1}-\mathbf{F} \mathbf{x}_{l}-\mathbf{s}_{l}\Arrowvert_{\mathbf{Q}_{l}^{-1}}^{2}    
 + \sum_{l=k}^{k+L-1}\lambda \Arrowvert \mathbf{D} \mathbf{s}_{l} \Arrowvert_{1}  \right\rbrace 
 \end{split}
\end{equation}
where $\Arrowvert \mathbf{x} \Arrowvert _{\mathbf{M}}^{2}=\mathbf{x}^{T}\mathbf{M}\mathbf{x}$, $\hat{\mathbf{x}}=[\hat{\mathbf{x}}_{1}, \ldots, \hat{\mathbf{x}}_{L} ]^{T}$ are the estimated states, $\hat{\mathbf{s}}=[\hat{\mathbf{s}}_{1}, \ldots, \hat{\mathbf{s}}_{L} ]^{T}$ are the estimated attacks, $\lambda$ is a regularization coefficient, and $\mathbf{D}$ is an $L \times 2L$  total variation  matrix  which forms the variation of the signal over time as~\cite{TV}
\vspace{-5pt}
\begin{equation}
\small
\mathbf{D}= \begin{pmatrix}
-1 & 0 & 1& 0&\ldots & 0 \\
0 & -1 & 0 & 1 & \ldots& 0 \\
\vdots & \vdots & \ddots & \ddots & \ddots & \vdots \\
0 & \ldots & 0 & -1 & 0 & 1 
\end{pmatrix}.
\end{equation}
The first term is the weighted residuals in the measurement equation, and the second term  is the weighted residuals of the state equation. The last regularization term promotes sparsity over the total variation of the estimated attack. 

\par In \eqref{proposedapproach}, the clock bias and clock drift are estimated jointly with the attack. Here, the model of the two introduced attacks should be considered. In Type I attack, a step attack is applied over the pseudoranges. The solution to the clock bias equivalently experiences a step at the attack time. The term $\Arrowvert \mathbf{D} \mathbf{s}_{l} \Arrowvert_{1}= \sum_{l=k+1}^{k+L-1} \left[  \arrowvert s_{b}[l] - s_{b}[l-1]\arrowvert+\arrowvert s_{\dot{b}}[l] - s_{\dot{b}}[l-1]\arrowvert \right] $  indicates a rise as it tracks the significant differences between two subsequent time instants. If the magnitude of   the estimated attack in two adjacent times does not change significantly, the total variation of the attack is close to zero. Otherwise, in the presence of an attack, the total variation of the attack includes a spike at the attack time. 

\par  In Type II attack, the total variation of the attack does not show significant changes as the attack magnitude is small at the beginning and the sparsity is not evident initially.  Although we explained why it is meaningful to expect only few nonzero entries in the total variation of the attacks in general, this is not a necessary condition for capturing the attacks during  initial small total variation magnitudes. This means that  explicit modeling of the attacks in \eqref{statemodel} and estimation through  \eqref{proposedapproach} does not require the attacks to exhibit sparsity over the total variation.  Furthermore, when the bias and bias drift are corrected using the estimated attack (we will provide one mechanism in Section IV-C), sparsity over the total variation appears for subsequent time instants.  In these time instants, the attack appears to be more prominent, and in effect,  the low dynamic behavior of the attack is magnified, a fact that facilitates the attack detection and will also be verified numerically. This effect is a direct consequence of \eqref{proposedapproach} and the correction scheme discussed in the next section.
\par The optimization problem of \eqref{proposedapproach} boils down to solving a simple quadratic program.  Specifically, the epigraph trick in convex optimization can be used to transform the $\ell_1$-norm into linear constraints~\cite{boyd_cvxbook}. The observation window $L$ slides for a lag time $T_{\mathrm{lag}} < L$, which can be set to $T_{\mathrm{lag}}=1$ for real-time operation. The next section details the sliding window operation of the algorithm, and  elaborates on how to use the solution of \eqref{proposedapproach} in order to provide corrected bias and drift. 
\vspace{-10pt}
\subsection{State Correction}
\label{State Correction}
In observation window of length $L$, the estimated attack $\hat{\mathbf{s}}$ is used to compensate the impact of the attack on the clock bias, clock drift, and measurements.
\par Revisiting the attack model in \eqref{statemodel}, the bias at time $l+1$ depends on the clock bias and clock drift at time $l$. This dependence  successively traces back to the initial time. Therefore, any attack on the bias that occurred in the past is accumulated through  time. A similar observation is valid for the clock drift. The clock bias at time $l$ is therefore contaminated by the cumulative effect of the attack on both the clock bias and clock drift in the previous times. The correction method takes into account the previously mentioned effect and modifies the bias and drift by subtracting the cumulative outcome of the clock bias and drift attacks as follows:
\vspace{-5pt}
\begin{equation}
\label{modificationmodel}
\begin{split}
\begin{pmatrix}
c\tilde{b}_{\mathrm{u}}[l]\\ \tilde{\boldsymbol \rho}[l]
\end{pmatrix}&= \begin{pmatrix}
c\hat {b}_{\mathrm{u}}[l] \\
\boldsymbol \rho[l]
\end{pmatrix} -\left( \sum_{l^{\prime}=k}^{l}\hat{s}_{b}[l^{\prime}]-\sum_{l^{\prime}=k}^{l-1}\hat{s}_{\dot{b}}[l^{\prime}] \Delta t \right)\mathbf{1}  \\
\begin{pmatrix}
c\tilde{\dot{b}}_{\mathrm{u}}[l] \\ \tilde{\dot{\boldsymbol \rho}}[l]
\end{pmatrix}&= \begin{pmatrix}
c\hat {\dot{b}}_{\mathrm{u}}[l] \\ \dot{\boldsymbol \rho}[l]
\end{pmatrix} -\left( \sum_{l^{\prime}=k}^{l}\hat{s}_{\dot{b}}[l^{\prime}] \right)\mathbf{1} 
\end{split}
\end{equation}
where $\tilde{b}_{\mathrm{u}}$ and $\tilde{\dot{b}}_{\mathrm{u}}$ are respectively the corrected clock bias and clock drift, $\tilde{\boldsymbol \rho}$ and $\tilde{\dot{\boldsymbol \rho}}$ are respectively the corrected pseudorange and pseudorange rates, and $\mathbf{1}$ is an all one vector of length $N+1$. In \eqref{modificationmodel}, $  l = 1,\ldots , L $ for the first observation window ($k=1$) and  $k+L-T_{\mathrm{lag}} \le l  \le k+L-1$ for the observation windows afterwards. This ensures that the measurements and states are not doubly corrected. The corrected measurements are used for  solving \eqref{proposedapproach} for the next observation window.
\par The overall attack detection and modification procedure is illustrated in Algorithm \ref{algorithm}. After the receiver collects $L$ measurements, problem \eqref{proposedapproach} is solved. Based on the estimated attack, the clock bias and clock drift are cleaned using \eqref{modificationmodel}. This process is repeated for a sliding window and only the clock bias and drift of the time instants that have not been cleaned previously are corrected. In another words, there is no duplication of modification over the states.
\par The proposed technique boils down to solving a simple quadratic program with only few variables and can thus be performed in real time. For example, efficient implementations of quadratic programming solvers are readily available in low-level programming languages. The implementation of this technique in GPS receivers and electronic devices is thus straightforward and does not necessitate creating new libraries.

\section{Numerical Results}
\label{Numerical Results}
We first describe the data collection device and then assess three representative detection schemes in the literature that fail to detect the TSA attacks. These attacks mislead the clock bias and clock drift, while maintaining correct location and velocity estimates. The performance of our detection and modification technique over these attacks is illustrated afterwards.
\begin{algorithm}[t!]
   \caption{: TSA Rejection and Mitigation (TSARM)}
   \label{algorithm}
   \begin{algorithmic}[1]
   \State Set $k=1$
   \While {True}
   \State Batch $\mathbf{y}_{l} \ \forall l = k,\ldots , k+L-1 $
   \State     Construct $ \mathbf{H},\mathbf{c}_{l}, \mathbf{F} \ \ \forall l = k,\ldots , k+L-1 $
   \State    Compute $\mathbf{Q}_{l}$ and $\mathbf{R}_{l}$ (details provided in Section V) 
   \State    Estimate $\hat{\mathbf{x}},\hat{\mathbf{s}}  $ via \eqref{proposedapproach}
   \State    Assign $c\hat {b}_{\mathrm{u}}[l]=\hat{\mathbf{x}}[m], m=2l-1  $ and $c\hat {\dot{b}}_{\mathrm{u}}[l]=\hat{\mathbf{x}}[m], m=2l $ $\ \ \forall l = k,\ldots , k+L-1$
   \State    Assign $\hat {s}_{b}[l]=\hat{\mathbf{s}}[m], m=2l-1  $ and $\hat{s}_{\dot{b}}[l]=\hat{\mathbf{s}}[m], m=2l $ $\ \forall l = k,\ldots , k+L-1$
   \State    Modify $\hat {b}_{\mathrm{u}}[l]$, $\hat {\dot{b}}_{\mathrm{u}}[l]$, $\boldsymbol \rho[l]$ and  $\dot{\boldsymbol \rho}[l]$ via \eqref{modificationmodel} $ \ \forall l = 1,\ldots , L $ for the first window and  $k+L-T_{\mathrm{lag}}  \le l  \le k+L-1$ for the windows afterwards
   \State Set $\mathbf{y}_{l} = \begin{pmatrix}
   \tilde{\boldsymbol \rho}[l] \\ \tilde{\dot{\boldsymbol \rho}}[l]
   \end{pmatrix}\ \forall l = k,\ldots , k+L-1 $
   \State Output $t_{R}^{\mathrm{UTC}}[l]= t_{R}[l]-\tilde{b}_{\mathrm{u}}[l]-\Delta t _{UTC} \ \forall l = k,\ldots , k+L-1$ to the user for time stamping
   \State    Slide the observation window by setting $k=k+T_{\mathrm{lag}}$
   \EndWhile
   \end{algorithmic}
   \end{algorithm}
\vspace{-5pt}
\subsection{GPS Data Collection Device}
\label{subsec:gpsdevice}
A set of real GPS signals has been recorded with a Google Nexus 9 Tablet at the University of Texas at San Antonio on June, 1, 2017.\footnote{Our  data is available at: \url{https://github.com/Alikhalaj2006/UTSA_GPS_DATA.git} } The ground truth of the position is obtained through taking the median of the WLS position estimates for a stationary device. This device has been recently equipped with a GPS chipset that provides raw GPS measurements. An android application, called \texttt{GNSS Logger}, has been released along with the post-processing MATLAB codes by the Google Android location team~\cite{GoogleSource}. 
\par Of interest here are the two classes of the \texttt{ \small Android.location} package. The \texttt{\small GnssClock}\footnote{\label{note1}\url{https://developer.android.com/reference/android/location/GnssClock.html} (accessed Feb. 20, 2017).} provides the GPS receiver clock properties and the \texttt{\small GnssMeasurement}\footnote{\label{note2}\url{https://developer.android.com/reference/android/location/GnssMeasurement.html} (accessed Feb. 20, 2017).} provides the measurements from the GPS signals both with sub-nanosecond accuracies. To obtain the pseudorange measurements, the transmission time is subtracted from the time of reception. The function \texttt{\small getReceivedSvTimeNanos()} provides the transmission time of the signal  which is with respect to the current GPS week (Saturday-Sunday midnight). The signal reception time is available using the function \texttt{\small getTimeNanos()}. To translate the receiver's time  to the GPS time (and GPS time of week), the package provides the difference between the device clock time and GPS time through the function \texttt{ \small getFullBiasNanos()}.
\par The receiver clock's covariance matrix, $\mathbf{Q}_{l}$, is dependent on the statistics of the device clock oscillator. The following model is typically adopted:
\begin{equation}
\mathbf{Q}_{l} =  \begin{pmatrix}
c^{2} \sigma_{b}^{2} \Delta t +c^{2}\sigma_{\dot{b}}^{2}\frac{\Delta t ^{3}}{3} & c^{2}\sigma_{\dot{b}}^{2} \frac{\Delta t ^{2}}{2} \\
c^{2}\sigma_{\dot{b}}^{2} \frac{\Delta t ^{2}}{2} & c^{2}\sigma_{\dot{b}}^{2} \Delta t 
\end{pmatrix} 
\end{equation}
where $\sigma_{b}^{2}=\frac{h_{0}}{2}$ and $\sigma_{\dot{b}}^{2}=2\pi^{2}h_{-2}$; and we select $h_{0}=8 \times 10 ^{-19}$ and $h_{-2}=2 \times 10 ^{-20}$~\cite[Chap. 9]{ Brown}. For calculating  the measurement covariance matrix, $\mathbf{R}_{l}$, the uncertainty of the pseuodrange and pseudorange rates are used.  These uncertainties are available from the  device together with the respective measurements.\footnoteref{note2} In the experiments, we set $\lambda=5 \times 10 ^{-10}$, because the distance magnitudes are in tens of thousands of meters. The estimated clock bias and drift through EKF in normal operation is considered as the ground truth for the subsequent analysis. In what follows, reported times are local.
\vspace{-5pt}
\begin{figure}
	\centering \includegraphics[scale=0.45]{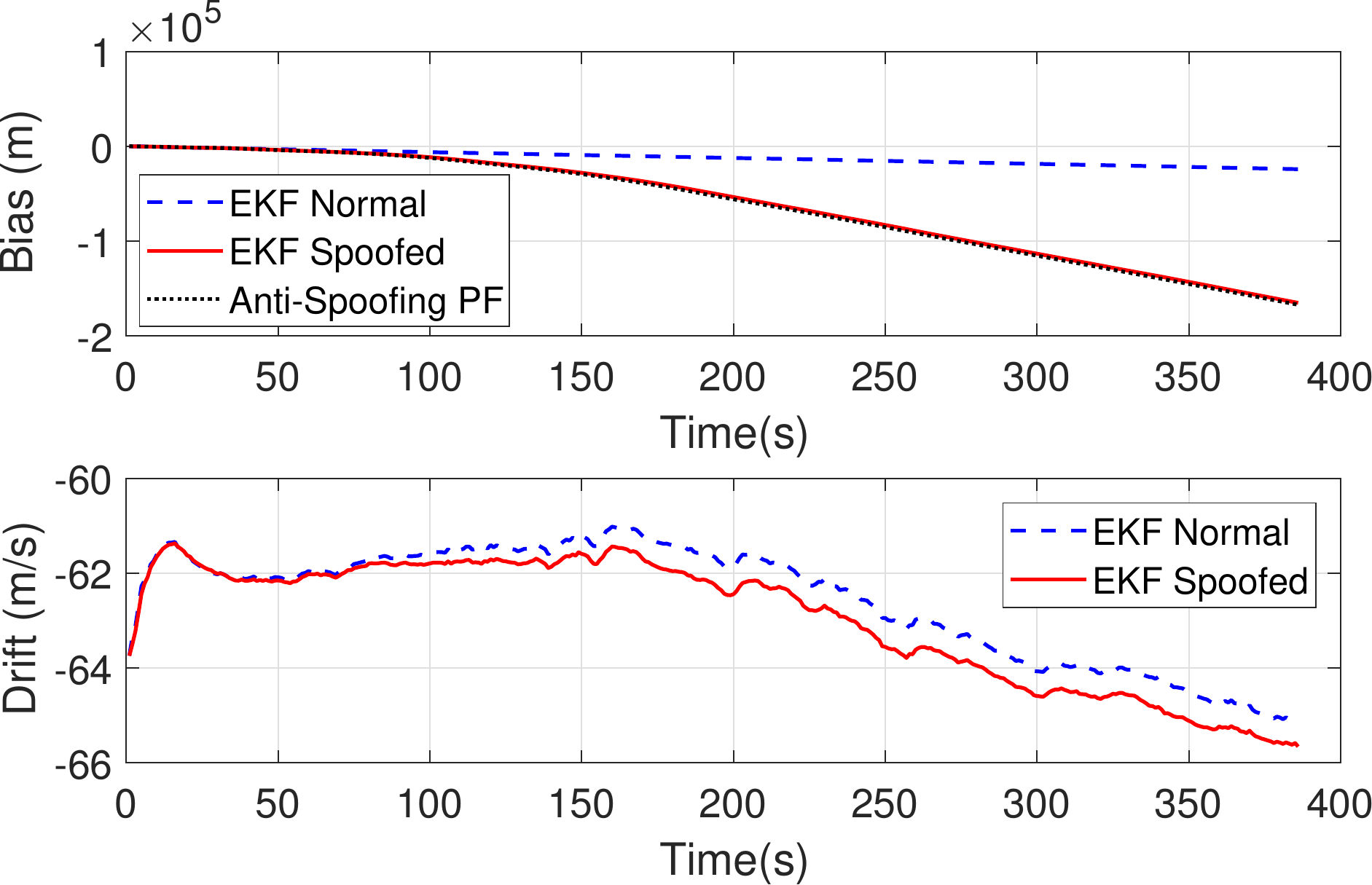}
	\caption{The effect of Type II attack on the EKF and the anti-spoofing particle filter \cite{ParticleFilter} on (a) clock bias and (b) clock drift. The attack started at  $t=30 \ \mathrm{s}$. Panel (b) does not include the drift.}
	\label{EKFfigure}
\end{figure}
\vspace{-5pt}
\subsection{Failure of Prior Work in Detecting Consistent Attacks}
This section demonstrates that three relevant approaches from Table \ref{table1} may fail to detect consistent attacks, that is, attacks where $s_{\rho}$ is the integral of  $s_{\dot{\rho}}$ in~\eqref{attack}.
\begin{figure}[t]
	\centering \includegraphics[scale=0.45]{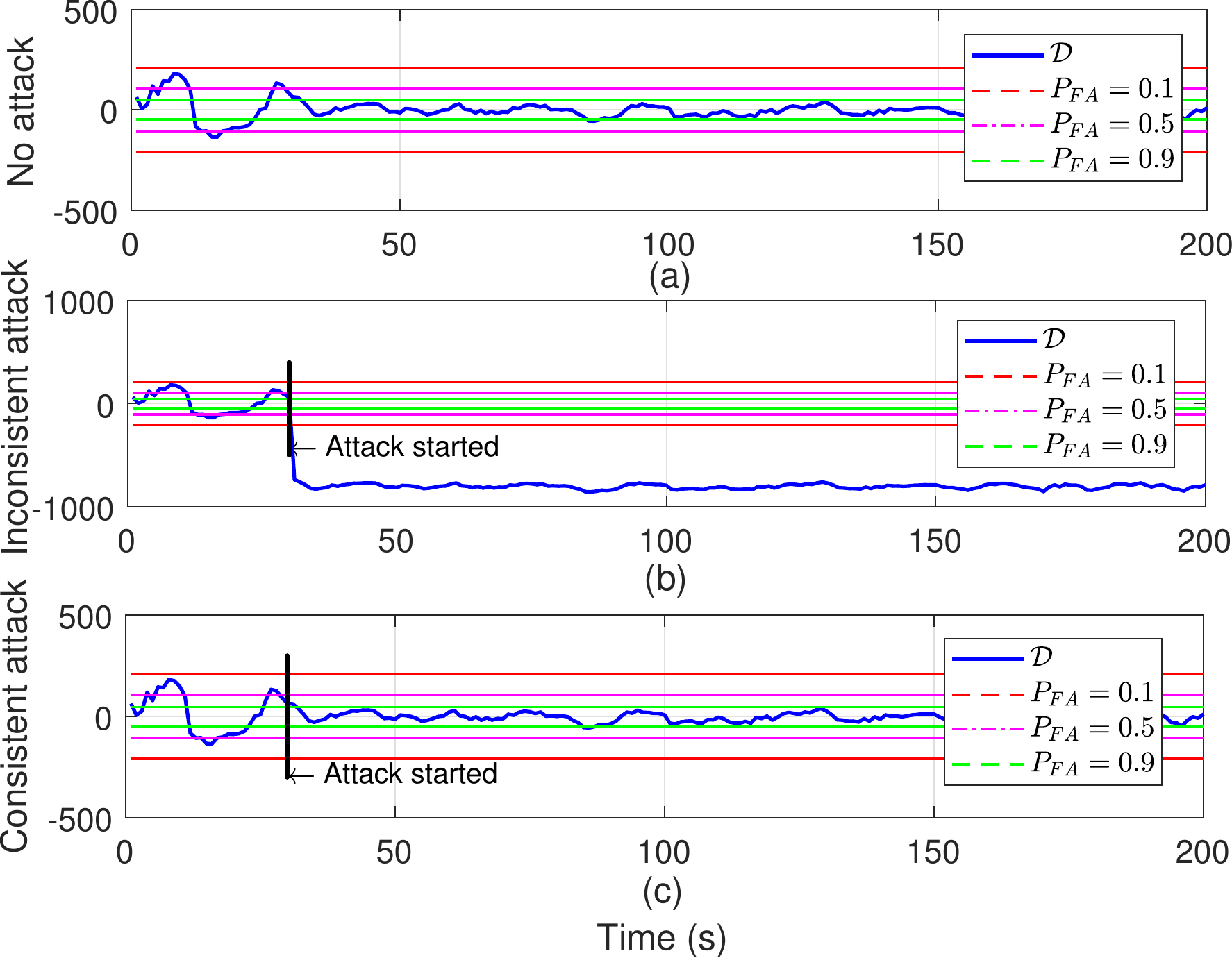}
	\caption{Performance of hypothesis testing based on statistic~\eqref{comparison}~\cite{r11} under Type I attack for different false alarm probabilities: (a) No attack, (b) inconsistent attack, (c) consistent attack. }
	\label{Comparison}
\end{figure}
\par The performance of the EKF and the anti-spoofing particle filter of~\cite{ParticleFilter} subject to a Type II attack is reported first.  The perturbations over GPS measurements are the same as in Fig.~\ref{SpooferII}  and are used as input to the EKF and the particle filter. The attack starts at $t=30 \ \mathrm{s}$.  Fig.~\ref{EKFfigure} depicts the effect of attack  on the clock bias and drift. The EKF on the dynamical model in \eqref{statemodel} and \eqref{measurementmodel} blindly follows the attack after a short  settling time. The anti-spoofing particle filter only estimates the clock bias and assumes the clock drift is known from WLS. Similarly to the EKF, the particle filter is not able to detect the consistent spoofing attack. The maximum  difference between the receiver estimated position obtained from the EKF on  \eqref{randomwalkmodel} under Type II attack and under normal operation is $x_{\mathrm{diff}}= 67 \ \mathrm{m}$, $y_{\mathrm{diff}}= 112 \ \mathrm{m}$, and $z_{\mathrm{diff}}= 71 \ \mathrm{m}$. The  position estimate has thus not been considerably altered by the attack.
\par The third approach to be evaluated has been proposed in~\cite{r11} and monitors the statistics of the receiver clock, as a typical spoofing detection technique~\cite{r7}. Considering that off-the-shelf GPS receivers compute the bias at regular $\Delta t$ intervals, a particular approach is to estimate the GPS time  after $k$ time epochs, and confirm that the time elapsed is indeed $k\Delta t$~\cite{r11}. To this end, the following statistic can be formulated:  
$
	\label{comparison}
	\small
	\mathcal{D}(k)=\left[ {t}_{R}^{GPS}(k)-t_{R}^{GPS}(1)-(k-1) \Delta t- \sum_{k^{\prime}=1}^{k} \hat{\dot{b}}[k^{\prime}] \Delta t \right] c .
$
The test statistic $\mathcal{D}$ is normally distributed with mean zero when there is no attack and may have nonzero mean depending on the attack, as will be demonstrated shortly. Its variance needs to be estimated from a few samples under normal operation.  The detection procedure relies on statistical hypothesis testing. For this, a false alarm probability, $P_{FA}$, is defined.  Each $P_{FA}$ corresponds to a threshold $\gamma$ to which $\mathcal{D}(k)$ is compared against~\cite[Chap. 6]{Detection}. If $\arrowvert \mathcal{D}(k) \arrowvert \ge \gamma $, the receiver is considered to be under attack.

The result of this method is shown in Fig.~\ref{Comparison} for different false alarm probabilities. Fig.~\ref{Comparison} (a) depicts  $\mathcal{D}(k)$ when the system is not under  attack. The time signature lies between the thresholds only for low false alarm probabilities. The system can detect the attack in case of an inconsistent Type I attack, in which $s_{\rho}$, is not the integration of perturbations over pseudorange rates,  $s_{\dot{\rho}}$, and only pseudoranges are attacked. Fig.~\ref{Comparison} (b) shows that the attack is detected right away. However, for smart attacks, where the spoofer maintains the consistency between the pseudorange and pseudorange rates, Fig. \ref{Comparison} (c) illustrates that the signature $\mathcal{D}(k)$ fails to detect the attack. This example shows that the statistical behavior of the clock can remain untouched under smart spoofing attacks. In addition, even if an attack is detected, the previous methods cannot provide an estimate of the attack.
\vspace{-7pt}
\subsection{Spoofing Detection on Type I Attack}
\par Fig. \ref{StepAttack} shows the result of solving~\eqref{proposedapproach} using the GPS measurements perturbed by the Type I attack of Fig. \ref{SpooferI}. The spoofer has the capability to attack the signal in a very short time so that the clock bias experiences a jump at $t=\mathrm{30 \ s}$. The estimated total variation of bias attack renders a spike right at the attack time.  The modification procedure of~\eqref{modificationmodel} corrects the clock bias using the estimated attack.
\vspace{-5pt}
\begin{figure}
	\centering \includegraphics[scale=0.45]{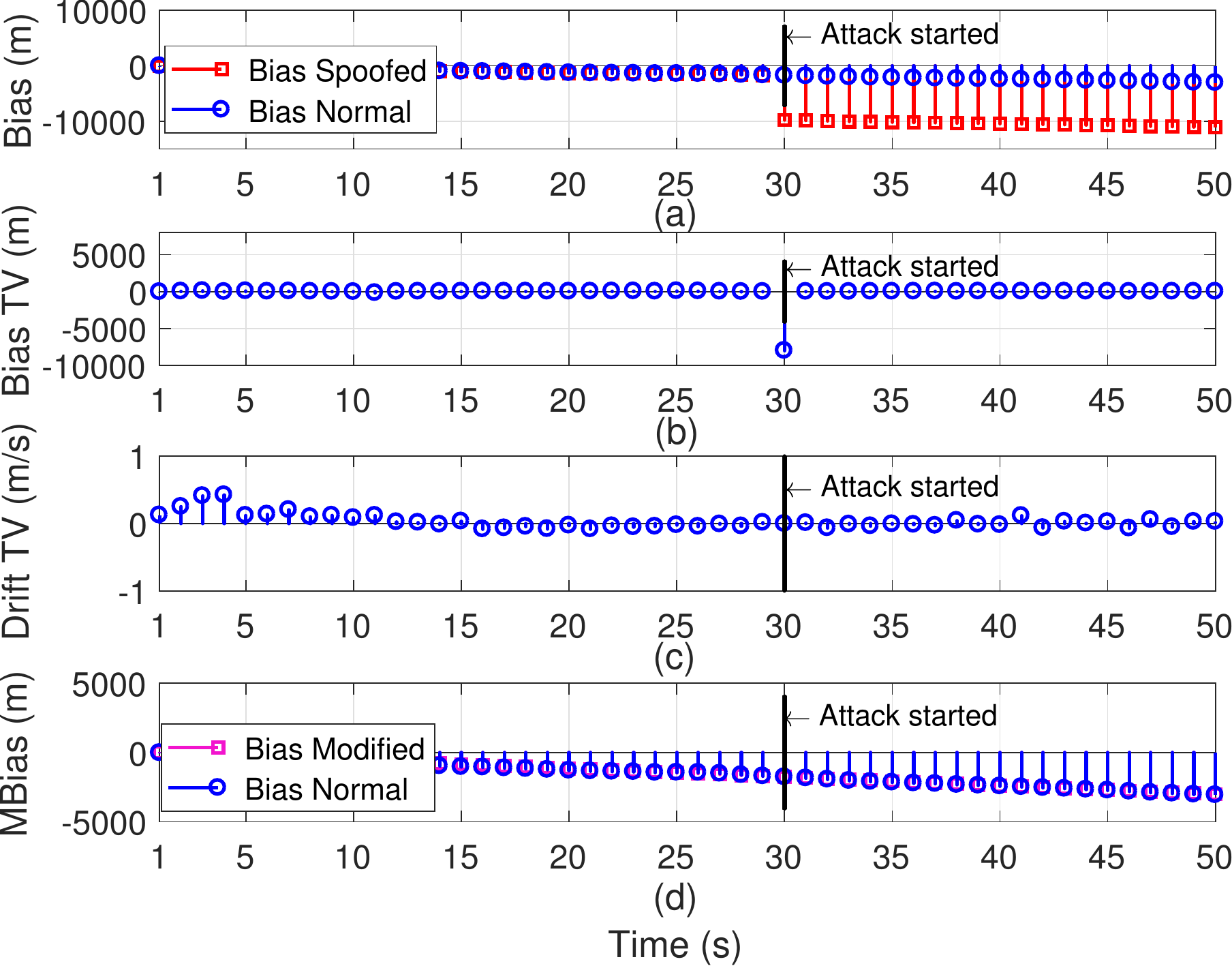}
		\caption{The result of attack detection and modification over a Type I attack that started at $t=30 \ \mathrm{s}$. From top to bottom: (a) Normal clock bias (blue) and spoofed bias (red), (b) total variation of the estimated bias attack  $\hat{\mathbf{s}}_{b}$, (c) total variation of the estimated drift attack $\hat{\mathbf{s}}_{\dot{b}}$, and (d) true bias (blue) and modified bias (magenta).}
	\label{StepAttack}
\end{figure}
\begin{figure}
	\centering \includegraphics[scale=0.45]{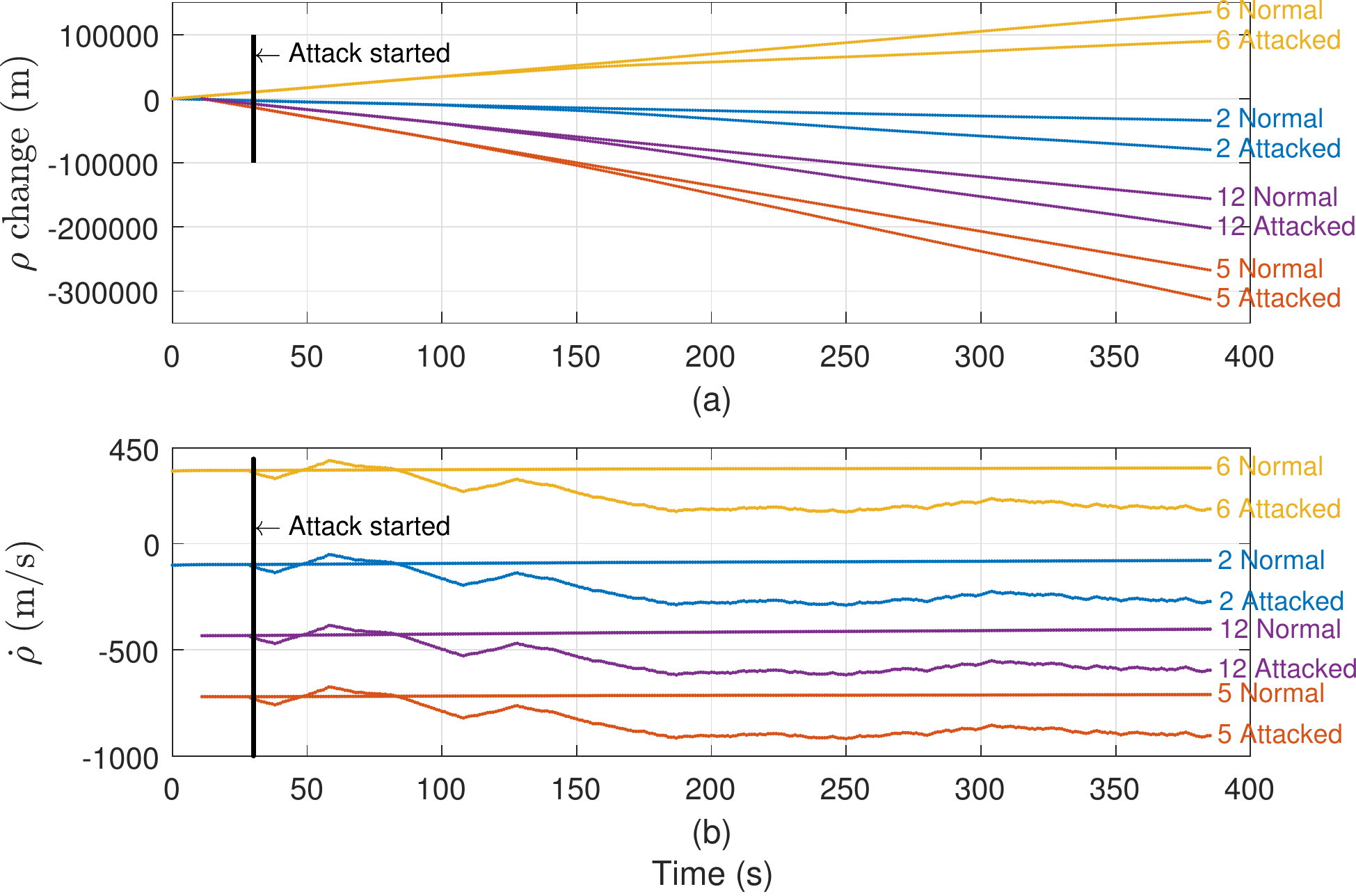}
		\caption{Comparison of (a) normal pseudorange change ($\rho(k)-\rho(1)$) and spoofed pseudoranges change ($\rho_{s}(k)-\rho_{s}(1)$), and (b) normal pseudorange rates ($\dot{\rho}$) and spoofed pseudorange rates ($\dot{\rho_{s}}$) under Type II attack for some of the visible satellites. The attack started at $t=30 \ \mathrm{s}$.} 
	\label{Prfigure}
\end{figure}
\begin{figure}
	\centering \includegraphics[scale=0.45]{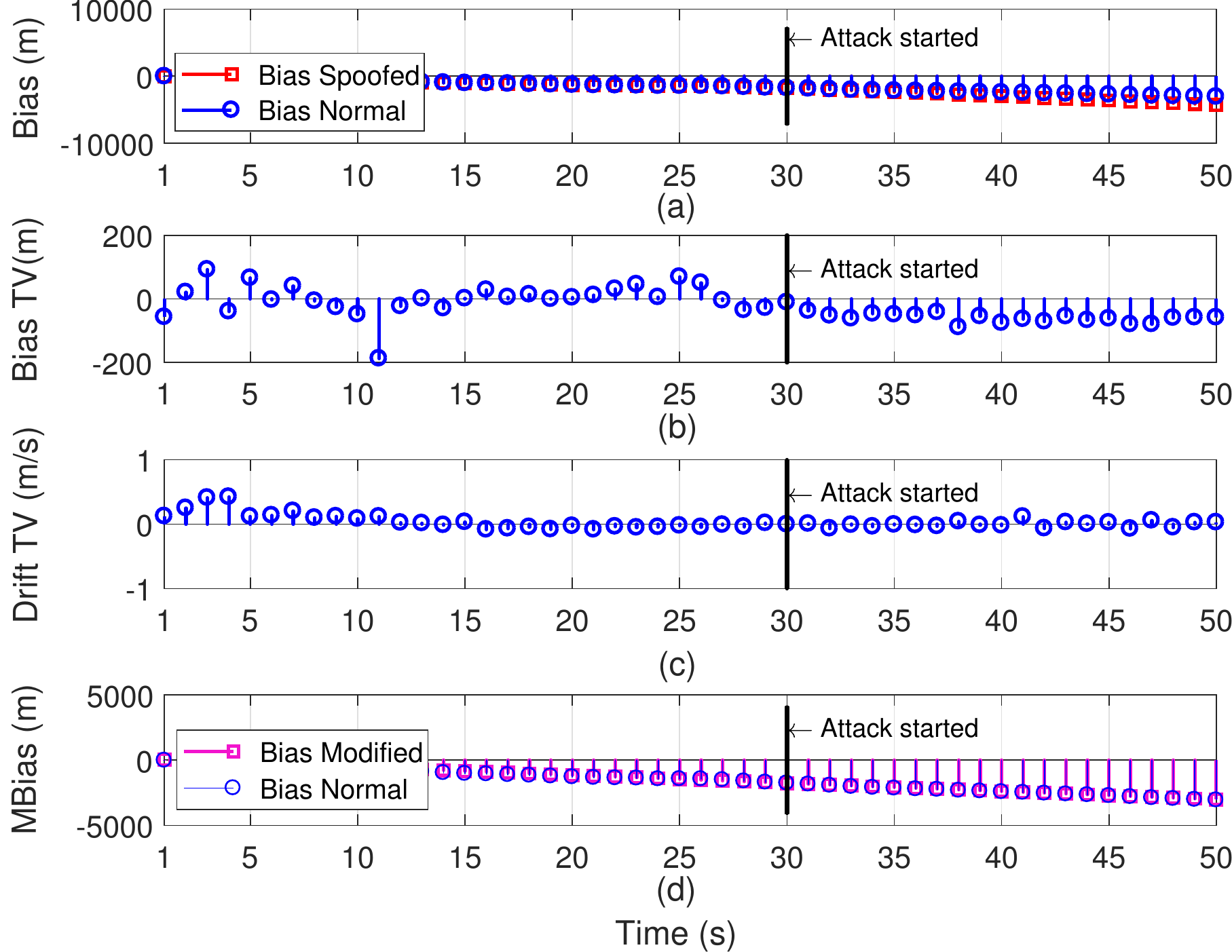}
		\caption{The result of attack detection and modification over Type II attack for $t=1 \ \mathrm{s}$ through $t=50 \ \mathrm{s}$. The attack started at $t=30 \ \mathrm{s}$. From top to bottom: (a) Normal clock bias (blue) and spoofed bias (red), (b) estimated bias attack  $\hat{\mathbf{s}}_{b}$, (c) total variation of the estimated bias attack, and (d) true bias (blue) and modified bias (magenta). }
	\label{First50Sec}
\end{figure}
\begin{figure}
	\centering \includegraphics[scale=0.45]{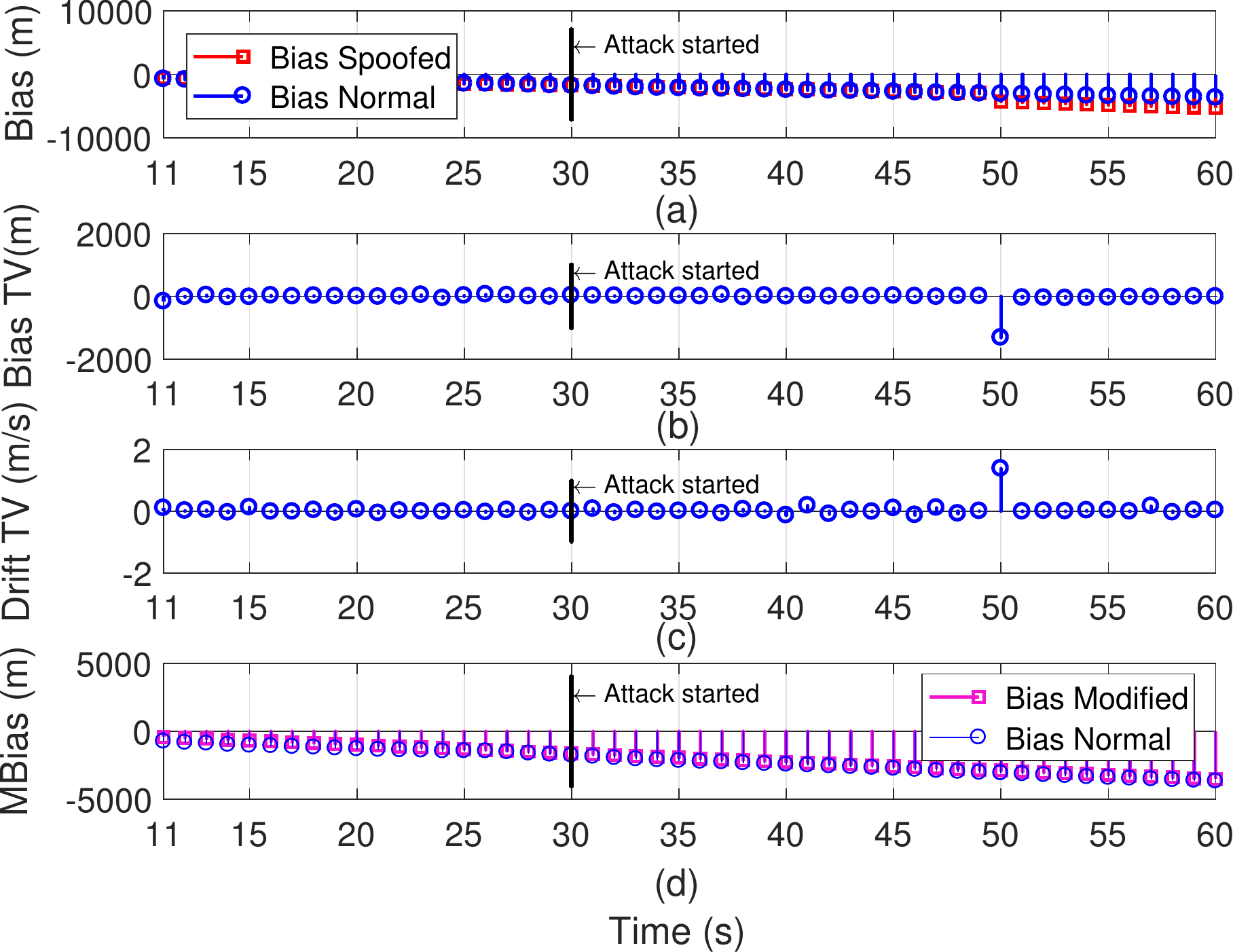}
		\caption{The result of attack detection and modification over Type II attack for $t=11 \ \mathrm{s}$ through $t=60 \ \mathrm{s}$.  From top to bottom: (a) Normal clock bias (blue) and spoofed bias (red), (b) estimated bias attack  $\hat{\mathbf{s}}_{b}$, (c) total variation of the estimated bias attack, and (d) true bias (blue) and modified bias (magenta). }
	\label{2070Sec}
\end{figure}
\vspace{-5pt}
\subsection{Spoofing Detection on Type II Attack}
The impact of Type II attack on the  pseudoranges and pseuodrange rates is shown in Fig.~\ref{Prfigure}. Specifically, Fig.~\ref{Prfigure} (a) illustrates the normal and spoofed pseudorange changes with respect to their initial value at $t=\mathrm{0 \ s}$ for some of the visible satellites in the receiver's view. Fig.~\ref{Prfigure} (b) depicts the corresponding pseudorange rates. The tag at the end of each line indicates the satellite ID and whether the pseudorange (or pseudorange rate) corresponds to normal operation or operation under attack. The spoofed pseudoranges diverge quadratically starting at $t=\mathrm{30 \ s}$ following the Type II attack.
\par For the Type II attack, Algorithm 1 is implemented for an sliding window with $L=\mathrm{50 \ s}$ with $T_{\mathrm{lag}}=\mathrm{10 \ s}$. Fig.~\ref{First50Sec} shows the attacked clock bias starting at $t=\mathrm{30 \ s}$. Since the attack magnitude is small at initial times of the spoofing, neither the estimated  attack $\hat{\mathbf{s}}_{b}$ nor the total variation do not show significant values.  The procedure of sliding window is to correct the current clock bias and clock drift for all the times that have not been modified previously. Hence, at the first run the estimates of the whole window are modified. Fig. \ref{2070Sec} shows the estimated attack and its corresponding total variation after one $T_{\mathrm{lag}}$.  As is obvious from the figure, the modification of the previous clock biases transforms the low dynamic behavior of the spoofer to a large jump at $t=\mathrm{50 \ s}$ which facilitates the detection of attack through the total variation component in \eqref{proposedapproach}. The clock bias and drift have been modified for the previous time instants and need to be cleaned only for $t=\mathrm{50 \ s}-\mathrm{60 \ s}$. 
\vspace{-5pt}
\subsection{Analysis of the Results}
Let $K$ be the total length of the observation time (in this experiment, $K=386$). The root mean square error (RMSE) is introduced:
$
	\small
	\mathsf{RMSE}= \frac{c}{K} \sqrt{\sum_{k=0}^{K-1}  (\tilde{b}_{u}[k]-\check{b}_{u}[k])^{2}}
$, which shows the average error between the clock bias that is output from the spoofing detection technique, $\tilde{b}_{u}$, and the estimated clock bias from EKF under normal operation, $\check{b}_{u}$, which is considered as the ground truth. Comparing the results of the estimated spoofed bias from the EKF and the normal bias shows that $\mathsf{RMSE}_{\mathrm{EKF}} = \mathrm{3882 \ m} $.  This error for the anti-spoofing particle filter is $\mathsf{RMSE}_{\mathrm{PF}} = \mathrm{ 3785 \ m} $. Having applied TSARM, the clock bias has been modified with a maximum error of  $\mathsf{RMSE}_{\mathrm{TSARM}} = \mathrm{258 \ m} $. Fig.~\ref{LTagfigure} illustrates the RMSE of TSARM for a range of values for the window size, $L$, and the lag time, $T_{lag}$. When the observation window is smaller, fewer measurements are used for state estimation. On the other hand, when $L$ exceeds $40 \ \mathrm{s}$, the number of states to be estimated grows although more measurements are employed for estimation. The numerical results illustrate that~\eqref{statemodel} models the clock bias and drift attacks effectively, which are subsequently estimated using \eqref{proposedapproach} and corrected through~\eqref{modificationmodel}. 
\begin{figure}
	\centering \includegraphics[scale=0.45]{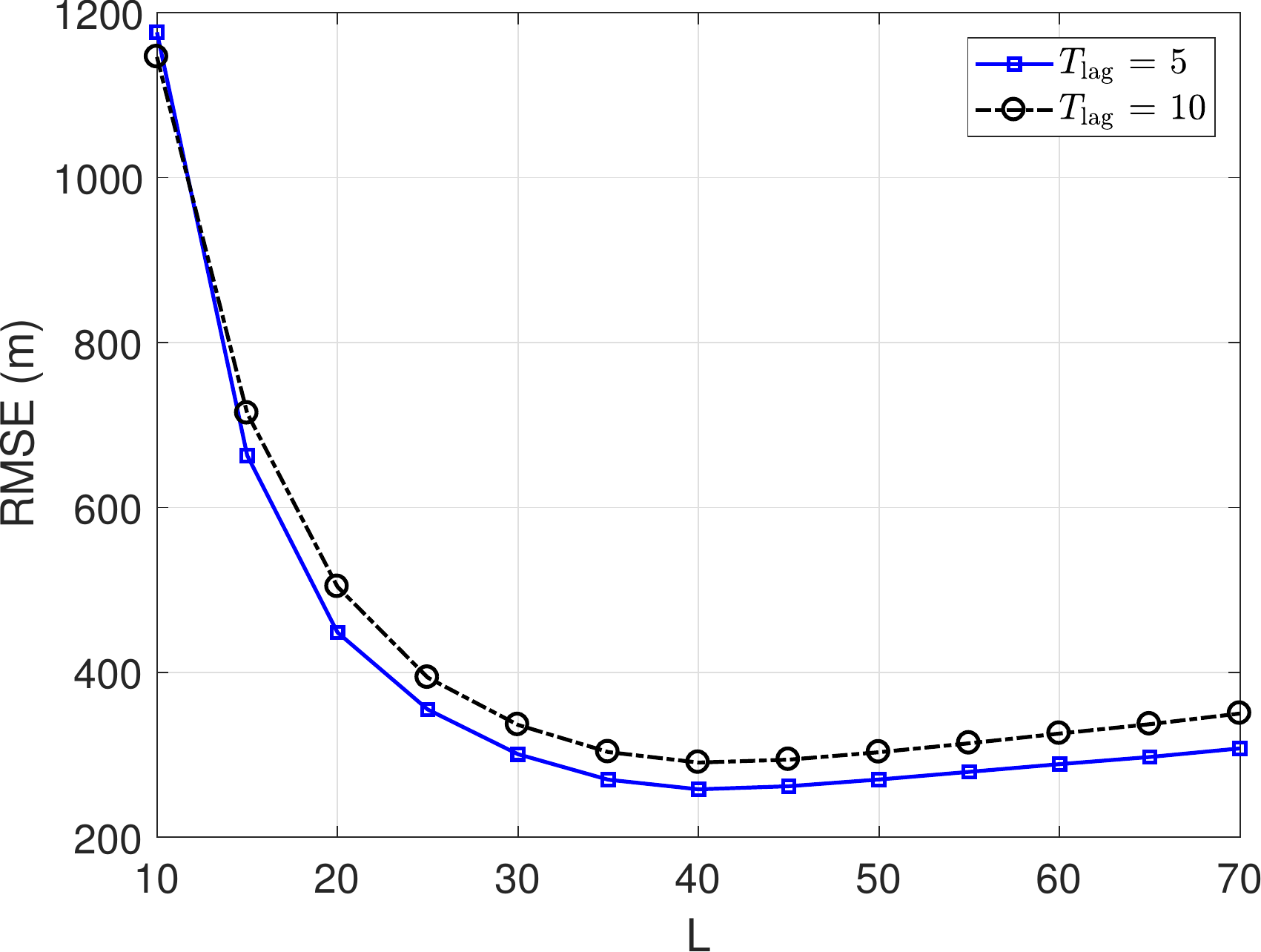}
	\caption{The RMSE of TSARM for various values of $L$ and $T_{lag}$. } 
	\label{LTagfigure}
\end{figure}
\vspace{-5pt}
\section{Concluding Remarks and Future Work}
\label{Conclusion}
This work discussed the research issue of time synchronization attacks on  devices that rely on GPS  for time tagging their measurements. Two principal types of attacks are discussed and a dynamical model that specifically models these attacks is introduced. The attack detection technique solves an  optimization problem to estimate the attacks on the clock bias and clock drift. The spoofer manipulated clock bias and drift are corrected using the estimated attacks. The proposed method detects the behavior of spoofer even if the measurements integrity is preserved. The numerical results demonstrate that the attack can be largely rejected, and the bias can be estimated  within $0.86 \ \mu \mathrm{s}$ of its true value, which lies within the standardized accuracy in PMU and CDMA applications. The proposed method can be  implemented for real-time operation.  

In the present work, the set of GPS signals are obtained from an actual GPS receiver in a real environment, but the attacks are simulated based on the characteristics of real spoofers reported in the literature. Experimentation on the behavior of the proposed detection and mitigation approach under real spoofing scenarios is the subject of future research.

% References

%\bibliographystyle{Bibliography/IEEEtranTIE}
%\bibliography{Bibliography/IEEEabrv,Bibliography/BIB_1x-TIE-2xxx}\ %IEEEabrv instead of IEEEfull

\bibliographystyle{IEEEtran}
%\bibliography{group_pubs,cif_proposal_refs,EAGER}
\bibliography{mybib}

{
	
	\begin{IEEEbiography}[{\includegraphics[width=1in,height=1.25in,clip,keepaspectratio]{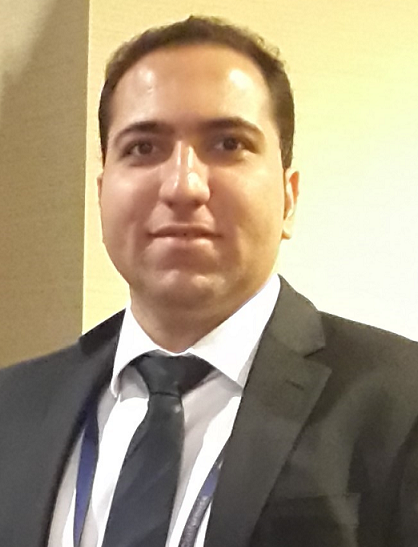}}]
		{Ali Khalajmehrabadi} (S'16) received the B.Sc.
		degree from the Babol Noshirvani University of
		Technology, Iran, in 2010, and the M.Sc. degree
		from University Technology Malaysia, Malaysia,
		in 2012, where he was awarded the Best Graduate
		Student Award. He is currently pursuing the Ph.D.
		degree with the Department of Electrical and
		Computer Engineering, University of Texas at
		San Antonio. His research interests include indoor
		localization and navigation systems, collaborative
		localization, and global navigation satellite system.
		He is a Student Member of the Institute of Navigation and the IEEE.
	\end{IEEEbiography}
	
	\vspace{-1 cm}
	\begin{IEEEbiography}[{\includegraphics[width=1in,height=1.25in,clip,keepaspectratio]{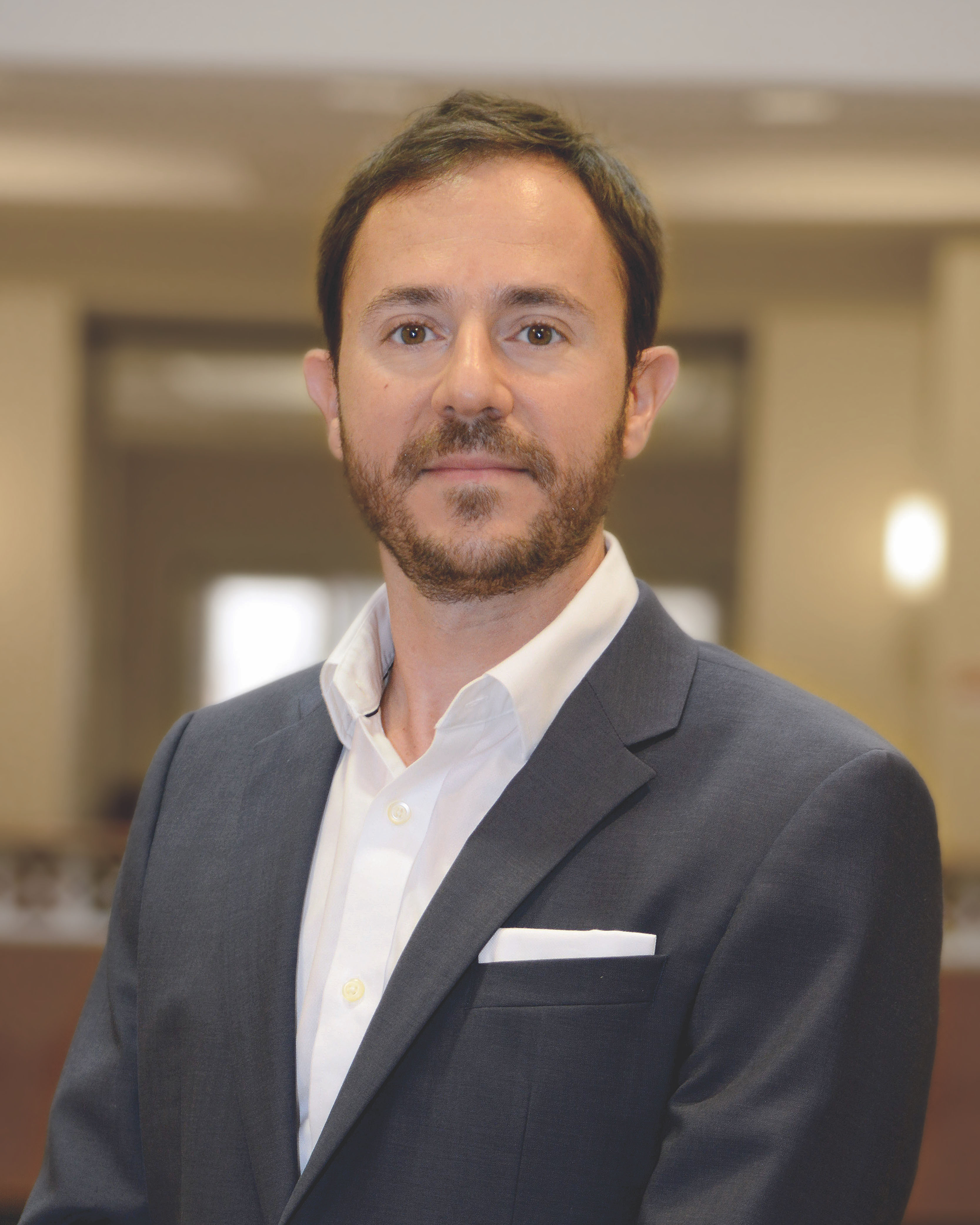}}]
		{Nikolaos Gatsis} (S'04-M'05) received the diploma
		(with Hons.) degree in electrical and computer engineering
		from the University of Patras, Greece, in
		2005, and the M.Sc. degree in electrical engineering
		and the Ph.D. degree in electrical engineering
		with minor in mathematics from the University of
		Minnesota, in 2010 and 2012, respectively. He is
				currently an Assistant Professor with the Department
				of Electrical and Computer Engineering, University
				of Texas at San Antonio. His research interests lie
				in the areas of smart power grids, communication
				networks, and cyberphysical systems, with an emphasis on optimal resource
				management and statistical signal processing. He has co-organized symposia in the area of smart grids in IEEE
				GlobalSIP 2015 and IEEE GlobalSIP 2016. He has also served as a co-guest editor for a special issue of the IEEE Journal on Selected Topics in Signal Processing on critical infrastructures. 
	\end{IEEEbiography}
	
	\vspace{-1 cm}
	\begin{IEEEbiography}[{\includegraphics[width=1in,height=1.25in,clip,keepaspectratio]{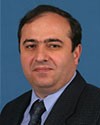}}]
		{David Akopian}  (M'02-SM'04) received the Ph.D.
		degree in electrical engineering in 1997. He is
		a Professor with the University of Texas at
		San Antonio. He was a Senior Research Engineer
		and a Specialist with Nokia Corporation from 1999
		to 2003. From 1993 to 1999, he was a Researcher
		and an Instructor with the Tampere University
		of Technology, Finland. He has authored and coauthored
		over 30 patents and 140 publications. His
		current research interests include digital signal processing
		algorithms for communication and navigation
		receivers, positioning, dedicated hardware architectures and platforms
		for software defined radio and communication technologies for healthcare
		applications. He served in organizing and program committees of many IEEE
		conferences and co-chairs annual SPIE Multimedia on Mobile Devices conferences.
		His research has been supported by the National Science Foundation,
		National Institutes of Health, USAF, U.S. Navy, and Texas foundations.
	\end{IEEEbiography}
	\vspace{-1cm}
	\begin{IEEEbiography}[{\includegraphics[width=1in,height=0.9in,clip]{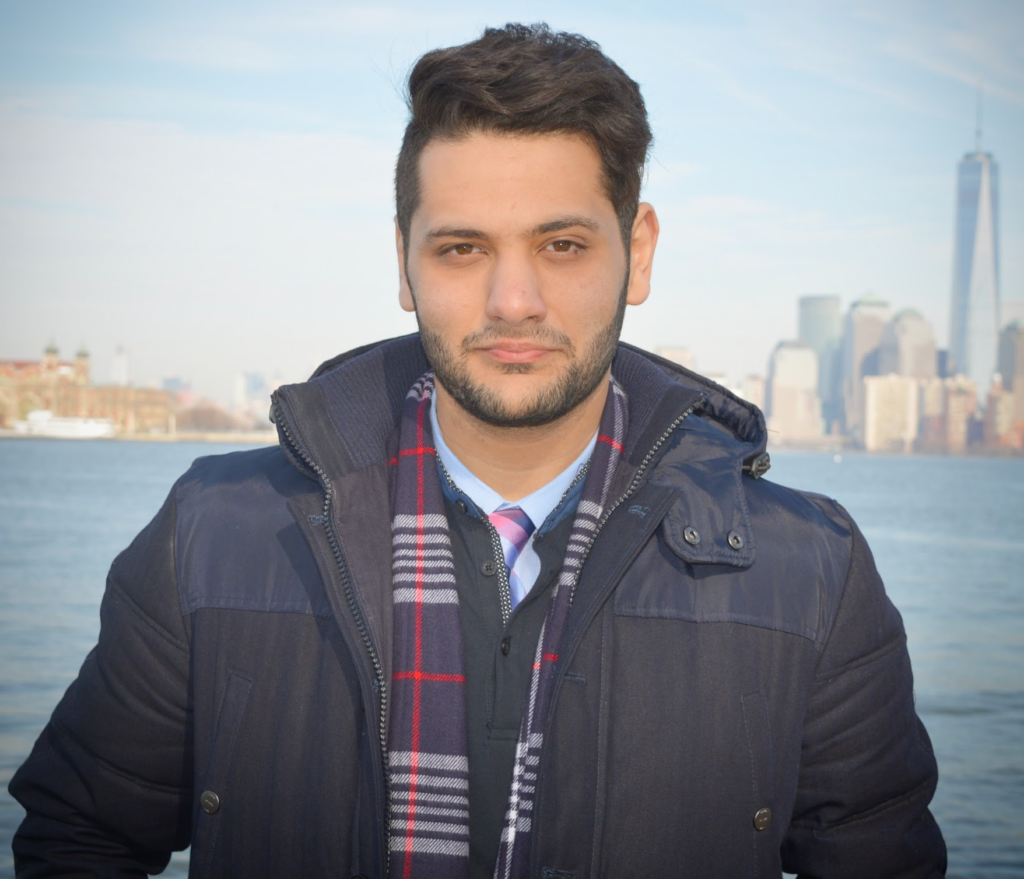}}]
		{Ahmad F. Taha}
		(S'07--M'15) received the B.E. and Ph.D. degrees in Electrical and Computer Engineering from the American University of Beirut, Lebanon in 2011 and Purdue University, West Lafayette, Indiana in 2015. In Summer 2010, Summer 2014, and Spring 2015 he was a visiting scholar at MIT, University of Toronto, and Argonne National Laboratory. Currently he is an assistant professor with the Department of Electrical and Computer Engineering at The University of Texas, San Antonio. Dr. Taha is interested in understanding how complex cyber-physical systems operate, behave, and \textit{misbehave}. His research focus includes optimization and control of power system, observer design and dynamic state estimation, and cyber-security.
	\end{IEEEbiography}
	
}

\end{document}